\documentclass[aps,superscriptaddress,twocolumn,floatfix]{revtex4}

\usepackage[dvips]{graphicx}
\usepackage{amssymb}
\usepackage{bm}

\begin{document}

\title{Simulation of structural phase transitions by metadynamics}

\author{R. Marto\v{n}\'ak}
\altaffiliation[Permanent address: ]{Department of Physics (FEI),
  Slovak University of Technology, Ilkovi\v{c}ova 3, 812 19 Bratislava,
  Slovakia}
\affiliation{Computational Science, Department of Chemistry and Applied
  Biosciences, ETH Zurich, USI Campus, Via Giuseppe Buffi 13, CH-6900
  Lugano, Switzerland}
\author{A. Laio}
\affiliation{Computational Science, Department of Chemistry and Applied
  Biosciences, ETH Zurich, USI Campus, Via Giuseppe Buffi 13, CH-6900
  Lugano, Switzerland}
\author{M. Bernasconi}
\affiliation{Dipartimento di Scienza dei Materiali and
Istituto Nazionale per la Fisica della Materia,
Universit$\grave{a}$ degli Studi di Milano-Bicocca,
Via Cozzi 53, I-20125 Milano, Italy}
\author{C. Ceriani}
\affiliation{Department of Chemical, Physical and Mathematical Science,
  University of Insubria at Como, Via Lucini 3, 22100 Como, Italy}
\author{P. Raiteri}
\affiliation{Computational Science, Department of Chemistry and Applied
  Biosciences, ETH Zurich, USI Campus, Via Giuseppe Buffi 13, CH-6900
  Lugano, Switzerland}
\author{M. Parrinello}
\affiliation{Computational Science, Department of Chemistry and Applied
  Biosciences, ETH Zurich, USI Campus, Via Giuseppe Buffi 13, CH-6900
  Lugano, Switzerland}

\date{\today}

\begin{abstract}
  We describe here in detail the recently introduced methodology for
  simulation of structural transitions in crystals. The applications of the
  new scheme are illustrated on various kinds of crystals and the
  advantages with respect to previous schemes are emphasized. The relevance
  of the new method for the problem of crystal structure prediction is also
  discussed.
\end{abstract}
\maketitle

\section{Introduction}
\label{sec:introduction}

The constant-pressure molecular dynamics (MD) method of Parrinello and
Rahman\cite{pr_md} enabled for the first time the study of structural phase
transitions in bulk solid crystals by computer simulation.  Starting from a
known initial structure it allowed the identification of possible
candidates for the new structure without any previous knowledge. It thus
achieved predictive power, in particular in combination with ab-initio
methods\cite{focher}, and has been successfully applied many times to a
variety of crystalline systems (for few selected applications, see
Refs.\cite{sissa1,sissa2,sissa3,sissa4,sissa5,sissa6,sissa7}). In the
practical use of the method, however, several problems arise related mainly
to the fact that structural transitions are often first order. This is by
necessity the case when the symmetries of the two crystal structures are
not in a group-subgroup relation. Experimentally, such transitions proceed
via nucleation of the new phase, which often starts on the surface or on
structural defects. For simulations of crystals, however, periodic boundary
conditions that eliminate surface are commonly used. The systems used in
simulations are typically relatively small and therefore contain no
structural defects.  The simulation setup therefore suppresses the
possibilities for a heterogeneous nucleation of the new phase. As a
consequence, the transformation of the system proceeds in a collective way,
involving all atoms, which results in a high barrier. This might cause a
metastability of the initial phase far beyond the thermodynamic transition
point and large hysteretic effects are frequently observed. Should there be
a substantial difference in volume between the phases, increasing pressure
favors the new phase with a smaller volume, due to the contribution of the
$PV$ term in the Gibbs potential. In order to observe the transition within
the accessible simulation time one often has to overpressurize the system
close to the point of mechanical instability\cite{kaxiras}.  Under such
conditions one or more intermediate phases may be
skipped\cite{focher,martins}, which reduces the predictive power of the
method. In other cases, where the volume difference between the phases is
not so pronounced, even overpressurization might not help to force the
transition to occur. For the above reasons, there is still a need for
developing specific methods aimed at simulating structural transitions in
crystals, given the great theoretical and practical relevance of the
closely related problem of crystal structure prediction.

In this paper we review recent progress in this field related to the
application of the new approach of Laio and Parrinello, called
metadynamics\cite{hills}. Rather than giving a ready-to-use recipe we would
like to highlight the possibilities offered by the new methodology which
allows the design of suitable algorithms for different kinds of systems. In
section \ref{sec:metadyn-using-simul} we review the generic algorithm
developed in Ref.\cite{prl} in which metadynamics is performed using the
simulation box as order parameter.  The use of the method will be
illustrated with examples of zeolite and benzene crystals. In section
\ref{sec:application-graphite} we discuss a different variant of the method
(Ref.\cite{zipoli}) suitable for systems where an internal order parameter
has to be used instead of the simulation box. This case is illustrated by
the example of graphite-to-diamond conversion. Finally, in the last part we
draw some conclusions and suggest possible directions for further
development.

\section{Metadynamics using a simulation box as order parameter}
\label{sec:metadyn-using-simul}

Since the Parrinello-Rahman method represents a generic constant-pressure
MD simulation method, the problems related to its application to structural
transitions originate in the lack of efficiency of standard MD in crossing
high barriers. During a standard equilibrium simulation the system explores
only a small part of its free energy surface, corresponding to thermal
fluctuations around a locally stable minimum. Consequently, a spontaneous
passage to another minimum separated by a barrier substantially larger than
the thermal energy $k_B T$ is extremely unlikely on the time scale of a
typical MD simulation, which is of the order of ns in the case of classical
force fields and ps in the case of ab-initio methods.

A new general approach for escaping free-energy minima and systematic
exploration of free-energy surfaces has recently been developed by Laio and
Parrinello\cite{hills}. It has been called \textit{metadynamics} and is
capable of dramatically speeding up the simulation of activated processes
involving barrier
crossing\cite{metad_app1,metad_app2,metad_app3,metad_app4}, including
first-order phase transitions. The general algorithm has been adapted by
Marto\v{n}\'ak, Laio and Parrinello \cite{prl} for simulating structural
transitions in crystals.  We note that this method represents a conceptual
extension of the idea of constant-pressure simulation. Instead of the
latter, we perform a search for new minima by \textit{exploring} the
surface of the Gibbs potential. In order to reduce the complexity of the
problem, an order parameter or collective variable is needed that allows us
to discriminate between different crystalline structures. A natural one for
a crystal is the structure factor $S(\mathbf{k})$ which provides a unique
fingerprint of the spatial arrangement of the atoms in a periodic lattice.
The structure factor, however, is not a convenient order parameter to use
for our purpose, because it has high dimensionality (in principle
infinite).  Instead, we follow in this point the approach of Parrinello and
Rahman and use as order parameter directly the three supercell edges
$\mathbf{a},\mathbf{b},\mathbf{c}$, arranged as a $3\times3$ matrix
$\mathbf{h}=(\mathbf{a},\mathbf{b},\mathbf{c})$. For relatively small
systems, where the creation of defects is too costly, the box matrix
$\mathbf{h}$ is likely to be simply related to the unit cell $\mathbf{u}$
via relation $\mathbf{h}=\mathbf{u}\mathbf{m}$, where $\mathbf{m}$ is an
integer matrix.  The matrix $\mathbf{h}$ can therefore distinguish between
different unit cells and crystal structures. Out of the 9 independent
degrees of freedom of the matrix $\mathbf{h}$ only 6 determine the shape of
the box, while the remaining 3 are related to the global rotation of the
box. The latter is irrelevant and only complicates the analysis of the
results.  Therefore it is convenient to freeze it, thus reducing the number
of degrees of freedom to 6. In Ref.\cite{prl} we followed the idea of
Nos\'e and Klein \cite{klein} and used a symmetric matrix $\mathbf{h}$. An
equally good choice is to constrain $\mathbf{h}$ to an upper triangular
form and here we follow this way. Since our aim is to simulate a phase
transition at a pressure $P$ and a temperature $T$ we will explore the
Gibbs potential $\mathcal{G}(\mathbf{h})=\mathcal{F}(\mathbf{h})+PV$ as a
function of $\mathbf{h}$ where $\mathcal{F}(\mathbf{h})$ is the Helmholtz
free energy of the system at fixed box and $V=\det(\mathbf{h})$ is the
volume of the box.  The six non-zero components of $\mathbf{h}$ can be
conveniently arranged as $\mathbf{h} = (h_{11}, h_{22}, h_{33}, h_{12},
h_{13}, h_{23})$ and represent a 6-dimensional order parameter. This
distinguishes the different local minima of $\mathcal{G}$ corresponding at
pressure $P$ to stable or metastable crystal structures.

The metadynamics requires calculation of derivatives of the free energy
with respect to the order parameter.  In our case such a derivative has a
simple form
\begin{equation}
\label{eq:19}
 -\frac{\partial \mathcal{G}}{\partial h_{ij}} =  
V \left[ \mathbf{h}^{-1} (\mathbf{p} - P)\right]_{ji} \;,
\end{equation}
where $\mathbf{p}$ is the internal pressure tensor, which can be easily
evaluated in MD or Monte Carlo simulations at constant $\mathbf{h}$ from
the averaged microscopic virial tensor \cite{allen}.

The order parameter is now evolved according to a steepest-descent-like
discrete evolution with a stepping parameter $\delta h$
(\emph{metadynamics})
\begin{equation}
\mathbf{h}^{t+1}=\mathbf{h}^{t}+\delta h \frac{\bm{\phi}^{t}}{\left|
    \bm{\phi}^{t}\right|  } \; .
\label{eq:11}
\end{equation}
Here, the driving force $\bm{\phi}^{t} = - \frac{\partial G^{t} }{\partial
  \mathbf{h}}$ is derived from a history-dependent Gibbs potential
$\mathcal{G}^{t}$ where a Gaussian has been added to
$\mathcal{G}(\mathbf{h})$ at every point $\mathbf{h}^{t^{\prime }}$ already
visited in order to discourage it from being visited again. Hence we have
\begin{equation}
  \mathcal{G}^{t}(\mathbf{h}) = \mathcal{G}(\mathbf{h}) + \sum_{t^{\prime }
    < t} W
  e^{-\frac{|\mathbf{h}-\mathbf{h}^{t^{\prime }}|^{2}}{2\delta h^{2}}}
\label{eq:12}
\end{equation}
and the force $\bm{\phi}^{t}$ is thus a sum of a thermodynamical driving
force $\mathbf{F} = - \frac{\partial \mathcal{G} }{\partial \mathbf{h}}$
and the term $\mathbf{F_{g}}$ coming from a potential constructed as a
superposition of Gaussians. As time proceeds the history-dependent term in
Eq.(\ref{eq:12}) fills the initial well of the free-energy surface and the
system is driven out of the local minimum.

\begin{figure}[!htb]
\begin{center}
  \includegraphics*[width=8.5cm]{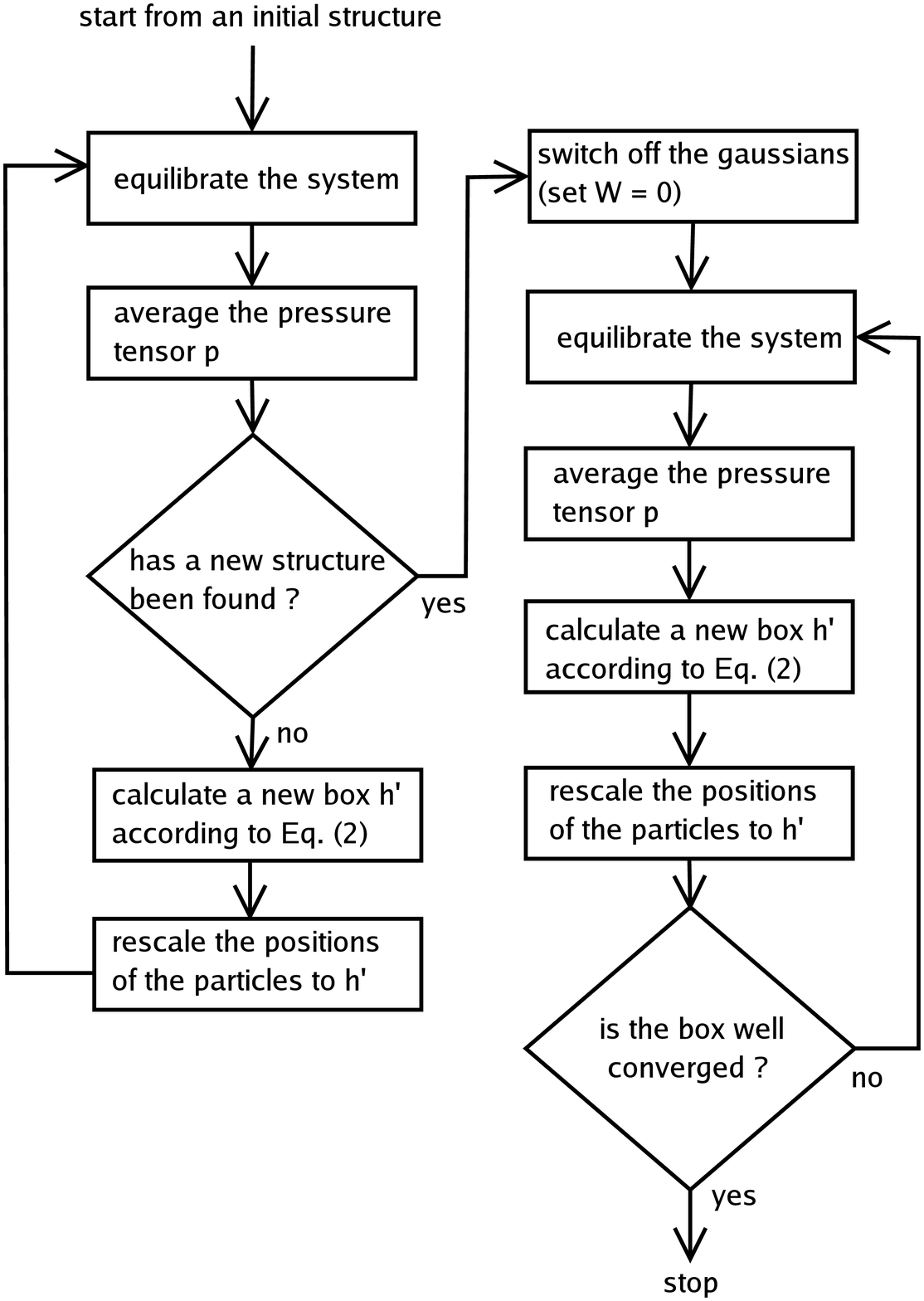}
\end{center}
\caption{Flowchart diagram of the metadynamics algorithm for seeking new
  crystal structures. \label{fig:flowchart} }
\end{figure}

The metadynamics algorithm described above can be implemented as follows
(Fig.\ref{fig:flowchart}).  We start from an equilibrated box $\mathbf{h}$
containing the initial structure at a given pressure $P$ and temperature
$T$ and evaluate the pressure tensor $\mathbf{p}$ in a constant
$\mathbf{h}$ MD run long enough to allow relaxation to equilibrium and
sufficient averaging of $\mathbf{p}$. The box $\mathbf{h}$ is then updated
using the forces (\ref{eq:19}) and metadynamics equations
(\ref{eq:11},\ref{eq:12}) to a new value $\mathbf{h}^{'}$. After the box is
modified the particle positions are rescaled in order to fit into the new
box using the relation $\vec{r}^{'} = \mathbf{h}^{'} \mathbf{h}^{-1}
\vec{r}$. In the case of molecular crystals we only scale the centers of
mass of the molecules but not the intramolecular degrees of freedom. As the
initial free energy well is gradually filled the box undergoes a sequence
of progressively larger deformations until a transition occurs and the
system enters into the basin of attraction of a new state. This can be
detected by monitoring the structure factor $S(\mathbf{k})$ and is often
apparent also on a visual inspection of the atomic configuration. At this
stage one can switch off the Gaussian term, so that the metadynamics
becomes purely steepest-descent-like and drives the system towards the
equilibrium state for the new structure. In this equilibrium state the
pressure will be equal to $P$.  Once the new structure is characterized one
can switch the Gaussians on again, thus filling the new minimum, and move
to other minima, if available.

The important point for the success of the method is the judicious choice
of the parameters $W$ and $\delta h$. In principle, these depend on the
$\mathcal{G}(\mathbf{h})$ landscape. The parameter $\delta h$ determines
the resolution in $\mathbf{h}$ and should be smaller than the typical size
of the well. A simple way to estimate the order of magnitude of $\delta h$
is to perform a short Parrinello-Rahman simulation and compute the
fluctuations of the box $\mathbf{h}$. However, we note that a very small
value of $\delta h$ is not likely to be useful. Since the volume of the
6-dimensional Gaussian is proportional to the sixth power of its size, a
small $\delta h$ means that many Gaussians would be needed to fill the
well, resulting in a long run.  In order to achieve the necessary energy
resolution, $W$ should be chosen as a fraction of the relevant energy
barriers. These are, however, not known at the beginning, since one does
not know the exact mechanism of the transition to the new phase.  A
practical guideline for the choice of $W$ and $\delta h$ can be based on
the requirement that the Gaussians should not be too ``sharp'', or in other
words, for an optimal filling the curvature of the Gaussians should be
smaller than that of the well (see Ref.\cite{hills} for a more general
discussion of the choice of $W$ and $\delta h$). This leads to the
condition $\frac{W}{{\delta h}^2} \leq K$ where $K$ is the smallest
eigenvalue of the $\mathcal{G}(\mathbf{h})$ Hessian at the minimum
$\mathbf{h_0}$. For a cubic system we can estimate $K$ from the approximate
expansion of $\mathcal{G}(\mathbf{h})$ around $\mathbf{h_0}$
\begin{equation}
\mathcal{G}(\Delta \mathbf{h}) \approx \mathcal{G}(\mathbf{h_0}) +
\frac{1}{2} V c (\frac{\Delta \mathbf{h}}{L})^2
\end{equation}
where $L$ is the cell edge and $c$ is of the order of magnitude of the
elastic constants. This provides an estimate $K \approx L c$ and a
condition
\begin{equation}
\label{eq:15}
\frac{W}{{\delta h}^2} \lesssim L c \; ;
\end{equation}
we stress here that the right-hand side of the inequality contains the box
size $L$ and therefore the choice of the parameters $W$ and $\delta h$ is
dependent on the size of the simulation box. In the examples that follow
the criterion (\ref{eq:15}) was found to provide a good working procedure.

We now discuss several aspects of the algorithm. First we note an important
difference with respect to the general method \cite{hills}. In principle,
metadynamics may allow the free energy profile to be recovered.  As shown
in Ref.\cite{hills}, the sum of the Gaussians in Eq.(\ref{eq:12}) converges
to $- \mathcal{G}(\mathbf{h})$ up to an additive constant, provided $W$ and
$\delta h$ are properly chosen, the system is confined to a finite region
of the order parameter space and several crossings between the minima
occur. Such information would, of course, be very valuable, since it would
allow an accurate determination of the transition pressure.  The latter is,
in general, difficult to calculate once the anharmonic effects start to
play a role and one cannot apply the quasiharmonic or self-consistent
quasiharmonic approximation. The recovering of the free energy, however,
implicitly assumes (similarly to thermodynamic integration) that the system
always evolves in a reversible way, and the free energy never decreases
discontinuously. This is generally not guaranteed in the case of
solid-solid transitions when box $\mathbf{h}$ is used as an order
parameter. In this case the metadynamics should be effectively regarded as
an ``engine'' which by inducing a progressively larger deformation of the
initial structure guarantees at some point the transition to a new one.
Whether such a transition proceeds in a reversible or irreversible way, as
well as the particular mechanism involved, is clearly very much system
dependent.  It can proceed via the creation of defects, such as grain
boundaries or stacking faults, or in a collective way, as when a mechanical
instability is reached by increasing the pressure in the Parrinello-Rahman
method. There is, however, a substantial difference related to the
symmetry. In the constant-pressure Parrinello-Rahman simulation the
internal stress tensor is on average equal to a prescribed value, which is
usually chosen as the hydrostatic pressure $P$. Such compression preserves
the initial symmetry of the system and unless the system is close to an
instability point, the symmetry is only slightly broken by instantaneous
thermal fluctuations. Metadynamics, however, is not a constant-pressure
simulation but rather a method aimed at exploring the free-energy surface.
Starting from an initial minimum, the exploration consists of applying
various symmetry-breaking deformations of the initial structure, which
clearly facilitates reaching a structure with a different symmetry. The
internal stress tensor during the exploration may become substantially
anisotropic and its fluctuations around the prescribed (hydrostatic) value
$P$ are much stronger than the thermal ones. For this reason, the role of
$P$ does not appear to be critical. As long as a given structure is at
least metastable at pressure $P$, it can in principle be found by our
algorithm; as discussed at the beginning of the previous section, the
phases are usually metastable in a broad pressure interval around the
equilibrium transition pressure $P_c$.  However, the value of $P$ affects
the number of steps necessary to reach the transition. If its value is too
low compared to $P_c$, a lot of Gaussians may be needed to fill the initial
well before a transition is observed.

A second remark should be made concerning the possibility that the system
might change box $\mathbf{h}$ without really changing the structure.  This
usually proceeds via plane sliding and is a manifestation of the fact that
one structure can be described by many equivalent choices of box
$\mathbf{h}$, known as modular invariance\cite{wentz}. This phenomenon
cannot be avoided completely as it is related to the very nature of the
order parameter $\mathbf{h}$.  Its frequent occurrence, however, together
with a failure of the algorithm to find a new structure, might be a
manifestation of either too wrong a pressure or the fact that box
$\mathbf{h}$ is not a good order parameter for a given system (see section
\ref{sec:metadyn-using-an}). According to our experience the choice of a
not too small value of $\delta h$, inducing substantial volume
fluctuations, helps to suppress the transitions to the same structure, as
these obviously conserve the volume. Too large a value might, however,
cause a transition to an amorphous structure; a suitable compromise has to
be found by trial and error.

Our original application to an atomic system, silicon crystal, is described
in the paper \cite{prl}; the method worked very well in that case. To
assess its applicability to a wider class of systems we also performed a
case study of two other crystals of a rather different kind.  These are
presented in the following subsections. Zeolite is an example of a crystal
having an extended network while benzene represents an organic molecular
crystal.

\subsection{Application to zeolite}
\label{sec:application-zeolite}

As our first example we provide here a summary of the main results obtained
by the application of the above method to reconstructive transitions on a
complex framework structure like a zeolite\cite{chiara}.

Zeolites are tectosilicates, with a structure formed by corner sharing
SiO$_{4}$ or AlO$_{4}$ tetrahedra and characterized by having cages or
channels able to accommodate alkaline or earth-alkaline cations and small
molecules (generally water)~\cite{galli,zeo}.  We have focused on the
Li-ABW zeolite (Li[AlSiO$_{4}$]$\cdot$H$_{2}$O), first synthesized by
Barrer and White in 1951~\cite{barrer}.  The framework structure has an
orthorhombic symmetry~\cite{norby} and is formed by directly connected
hexagonal rings sheets in the $bc$ plane.  Water molecules and Li cations
are located inside monodimensional 8 membered channels developing along the
$c$ axis.  If the temperature is raised in dry environment, the Li-ABW
zeolite undergoes two phase transitions.  A first displacive transition
occurs to the framework when dehydration starts, leading to a structure
known as anhydrous Li-ABW. At a temperature around 650$^{\circ}$C a second
reconstructive transition drives the system to a new structure, the
$\gamma$-eucryptite.  This transition is accompanied by a symmetry change,
leading to a monoclinic crystal.  The framework, formed by 8- and
4-membered rings of tetrahedral units in the anhydrous Li-ABW, transforms
in to a 6-membered rings structure. Because of the high stability of the
Al-O and Si-O bonds a Parrinello-Rahman approach is not able to reproduce
the reconstructive transition; in order to provide enough thermal energy to
observe a spontaneous breaking of some bonds, the temperature has to be
increased so much that only a collapse of the structure is observed.

The simulation of the anhydrous Li-ABW $\to \gamma$-eucryptite transition
has been carried out within the metadynamics approach employing a classical
potential, using the form and the parameters proposed in literature by Zirl
and Garofalini~\cite{garofalini}.  Details about the structure of the
potential and its adaptation for our purposes can be found in
Ref.\cite{chiara}.

\begin{figure}[!htb]
\begin{center}
  \includegraphics*[width=8.5cm]{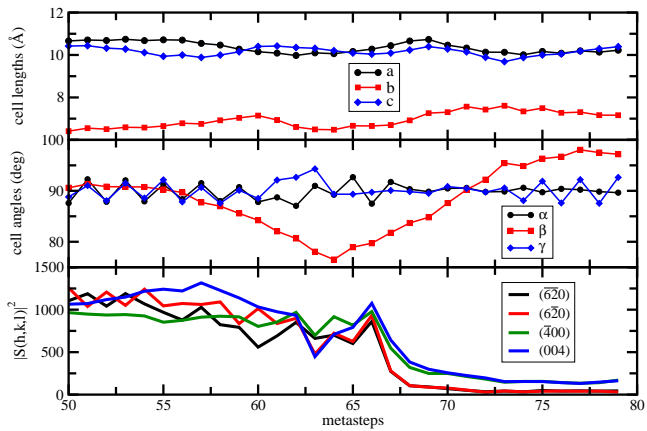}
\end{center}
\caption{Zeolite: time evolution of cell lengths, angles and selected
  strong peaks of the structure factor during the metadynamics
  simulation.\label{fig:abc}}
\end{figure}

\begin{figure}[!htb]
\begin{center}
  \includegraphics[width=8.5cm]{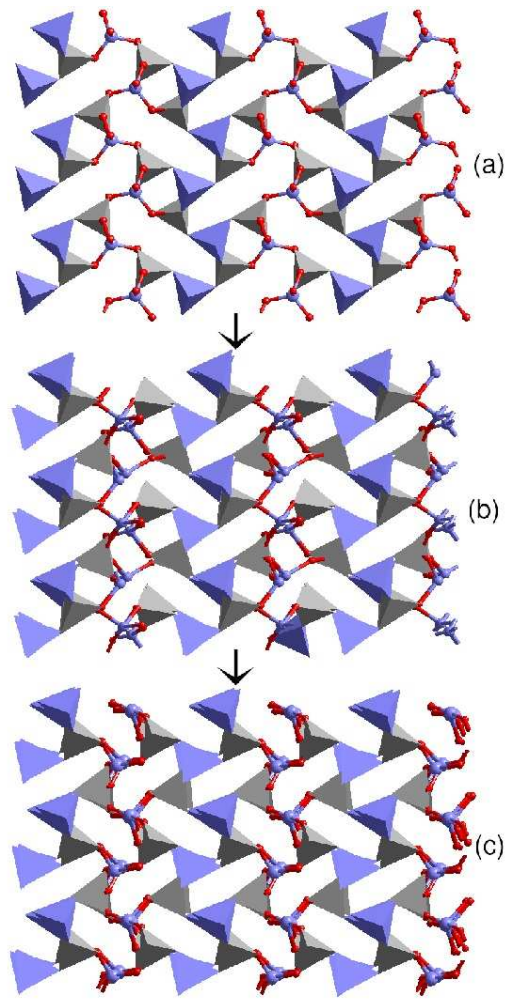}
\end{center}
\caption{Zeolite: three main steps involved in the transition path.
  The blue tetrahedra are the Al-centered ones, while the grey tetrahedra
  are those that are Si-centered.  The tetrahedra directly involved in the
  bond breaking are pictured in ball-and-stick representation, where the
  red spheres represent the oxygen atoms.  The Li cations inside the
  cavities are omitted for clarity.  (a) The anhydrous Li-ABW structure (8-
  and 4-membered rings), (b) an intermediate structure, formed during the
  first bond switching, (c) the final $\gamma$-eucryptite structure
  (6-membered rings).\label{fig:3steps}}
\end{figure}

The metadynamics run was performed at external pressure $P=0$ for a system
consisting of 56 atoms, using the parameters for the Gaussian term equal to
$W$ = 9.94\,kcal/mol and $\delta h$ = 0.4\,\AA. The total simulation time
for each NVT run was 10 ps with an equilibration of 2.5 ps. During the
first 67 steps of metadynamics the deformation of the simulation cell does
not lead to bond breaking, but only to displacive rearrangement of the
framework. After step 67 a reconstructive event is recognized by a sudden
drop of the strongest diffraction peaks of the original structure (see
Fig.\ref{fig:abc}). The transition from anhydrous Li-ABW to
$\gamma$-eucryptite is completed in three metadynamics steps and it
involves only processes of bond switching between some Al-O bonds
(Fig.\ref{fig:3steps}). In particular, half of the Al atoms of the
simulation cell move in a concerted way, leading to four bond switch, such
that the Li-ABW tetrahedral units are broken.  After few picoseconds, a
second bond switch is observed for each of the Al atoms already involved in
the first step, leading to the formation of new tetrahedral units with the
topology of the eucryptite. While the Al atoms involved in the bond
breaking move significantly, exploring three main positions, the other
atoms at the center of the unbroken tetrahedral units move very little from
their starting crystallographic positions, in spite of the fact that each
tetrahedron undergoes a rigid rotation in order to complete the transition.
The Li atoms also participate in the transformation with a displacement
from their starting position near to the Al involved in the bond switch.
The transition pathway observed with metadynamics is in good agreement with
the experimental observations reported in Ref.~\cite{norby}. A more
detailed description of the calculations and the transition path are given
in Ref.~\cite{chiara}.

The above results demonstrate the ability of metadynamics to uncover the
detailed microscopic mechanism of a phase transition in a complex crystal
structure. Another attractive feature of the procedure we outlined in the
previous section is that it allows a systematic exploration of the
different polymorphs in which a given system can exist. Starting from an
initial structure, several metadynamics runs can be performed at different
conditions of pressure and temperature until a transition is observed.  Any
classical potential providing a {\it qualitative} description for the
system can be employed, and performing the metadynamics with an accurate
model (e.g.~DFT) might be a waste of computer time.  In fact, the Gibbs
free energy difference between the candidate structures found by
metadynamics can be computed with the high level method {\it once the
  structures are known} at a relatively small computational price.

\begin{figure}[!htb]
 \begin{center}
   \includegraphics*[width=8.5cm]{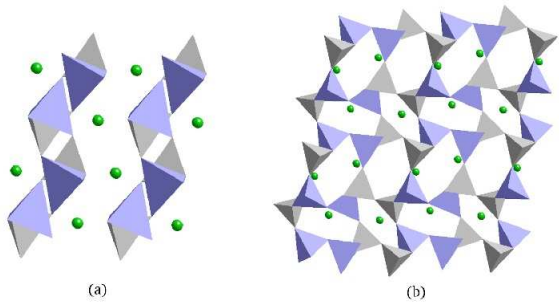}
 \end{center}
 \caption{Zeolite: two phases obtained by performing metadynamics at high
   external pressure. More details will be provided in a forthcoming
   publication\cite{chiara_new}. \label{fig:newphases} }
\end{figure}

For the system under investigation, we performed several metadynamics runs
starting from the Li-ABW structure in the pressure range between 10 and 100
kbar, obtaining seven new structures characterized by a different
connectivity of the atoms. To illustrate the excellent capability of the
technique to find new polymorphs we report here a brief description of two
of the structures we find (these results will be described in detail in a
separate publication \cite{chiara_new}).  The phase pictured in
Fig.\ref{fig:newphases} (a) is formed by ordered sheets, balanced by the
presence of Li cations in the intra-layers space.  Each layer is formed by
alternating Si-centered and Al-centered tetrahedra that form four-membered
rings.  The structure reported in Fig.\ref{fig:newphases} (b) has the same
connectivity of $\gamma$-eucryptite, but in this case there is no perfect
alternation of Si-centered and Al-centered tetrahedra, whereas there are
some Al-O-Al and Si-O-Si bridges.  The Al-O-Al bridge is normally not found
in zeolitic frameworks (L\"owenstein's rule\cite{lowenstein}) because it
shows larger stability for angles around 180$^{\circ}$ which are generally
not present in zeolites, but in conditions of high external pressure such
configuration can be stabilized.  For these two phases and the other five
new structures found by the metadynamics procedure we determined the {\it
  ab-initio} equation of state and the phonon spectrum in order to estimate
the entropic contribution to the free energy, thus obtaining a phase
diagram for the system at the DFT level and without any {\it a priori}
knowledge of the possible phases of the system\cite{chiara_new}.

\subsection{Application to benzene}
\label{sec:application-benzene}

The possibility of predicting all of the stable crystalline structures and
thus the physical and chemical properties for a given compound is extremely
important for pharmaceuticals\cite{pharma1,pharma2} and benzene is a good
case study to prove the applicability of our technique to organic
molecules. Benzene, though being a simple molecule has a quite complex
phase diagram.  Since the late 60's many authors studied solid benzene and
in the literature two different notations for its crystalline phases can be
found.  In this paper we follow the notation used by Thi\'{e}ry and
L\'{e}ger in Ref.~\cite{thiery}.  At room temperature and ambient pressure
benzene crystallizes in an orthorhombic structure, {\it{Pbca}}, which is
stable up to 1.4~GPa~\cite{thiery}.  Beyond this pressure a sluggish
transition to benzene II is observed~\cite{thiery,cansell,ciabini}.
Determining the crystalline structure of this phase was a big challenge and
only 10 years after its first observation theoretical calculations could
determine that it is orthorhombic and belongs to the {\it{P4$_3$2$_1$2}}
space group~\cite{eijck}. Benzene II is stable up to 4~GPa and then it
transforms into benzene III, which is monoclinic and belongs to the
{\it{P2$_1$/c}} space group~\cite{thiery,cansell,ciabini}.  Upon increasing
pressure two more phases are observed: benzene III', which is stable
between 11 and 24~GPa, and benzene IV, which is stable at even higher
pressure\cite{thiery,ciabini}.  These latter two structures are not well
characterized and there is still some debate on whether they are real
phases or not. In particular, benzene III' is supposed to be only a
modification of benzene III\cite{thiery} and benzene IV to be
polymorphic~\cite{cansell}.  Beyond these five, another phase has been
hypothesized: benzene I', which should be stable at low temperatures
($\sim$100~K)~\cite{thiery} and normal pressure.

In this work we focus our attention only on benzene I, II and III. These
are indeed the only three phases for which the structure is well understood
both by experiment and by theory. In a different publication we tackle the
problem of determining all stable structures of benzene ~\cite{jacs} and
show that the method can indeed identify the structure of all
experimentally proposed phases.

All our MD simulations were done using the GROMOS96 force
field\cite{gromos}. The molecule is fully flexible and the long range
electrostatic interaction is calculated with the Particle Mesh Ewald (PME)
summation. We note that the intermolecular potential can reproduce the
stability of all the three phases, though not their correct transition
pressures. At 0~K, benzene II is less stable than benzene III and, at
variance with the experiments, the common tangent construction suggested a
transition pressure of about 1.5~GPa for both the I to II and I to III
transformation\cite{thiery,cansell}.

We apply metadynamics to study the transitions among those phases.  During
the metadynamics both the equilibration and the averaging of the internal
stress tensor are performed by NVT runs of 1~ps, the external pressure $P$
is always fixed at 2~GPa and the temperature at 300~K. For what concerns
the parameters entering the time dependent potential, we tried a wide range
of values, however, good results are obtained only when the Gaussian width,
$\delta h$, is chosen between 2 and 3~\AA\ and the Gaussian height,
{\it{W}}, between 120 and 600 kcal/mol. Given the elastic constants of
benzene of about 6~GPa and the box vectors varying between about 10 and
50~\AA\ we notice that the working parameters fulfill the empirical
guideline given in section \ref{sec:metadyn-using-simul} (Eq.~\ref{eq:15}).
We experienced that the correct choice of $\delta h$ is extremely
important. In fact, if $\delta h < 2$~\AA\ during metadynamics run we only
observe a useless plane sliding, which is indeed a much less expensive
deformation than a genuine solid--solid phase transition. For molecular
solid this is even more easy than in covalent crystals because of the weak
nature of the intermolecular interactions. Moreover, if $\delta h >
3$~\AA\, the deformation of the cell was too large to be accommodated by
the molecules and amorphization is always observed. On the other hand,
choosing a too small value for $W$ causes a slow escaping from the energy
basin resulting in a waste of time.

\begin{figure}[!htb]
  \begin{center}
    \includegraphics*[width=8.5cm]{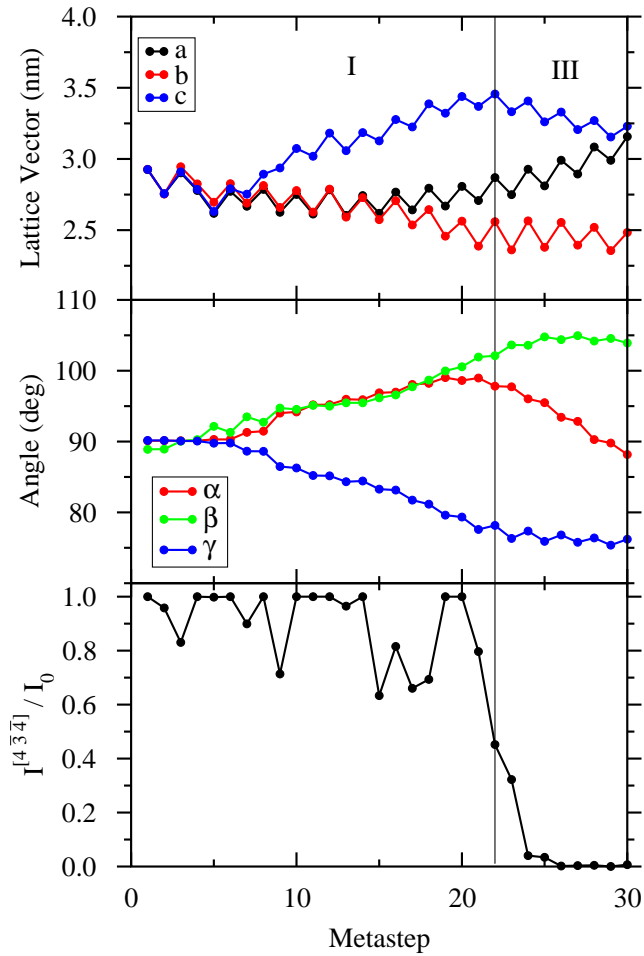}
  \end{center}
  \caption{Benzene: behavior of the box lengths and angles in a typical
    metadynamics run with 192 molecules in the simulation cell (top and
    middle panels). In the bottom panel we report the intensity of the
    $(4\overline{3}\overline{4})$ reflection relative to the strongest peak
    observed in that metastep. The sudden drop of the intensity clearly
    indicates the occurrence of a phase transition at the metastep 22.
  \label{fig:meta_1}}
\end{figure}

The structure factor provides a unique fingerprint of each crystalline
structure and can be compared with the experimental x-ray powder
diffraction pattern. Characteristic reflection peaks of a given structure
often vanish for other crystalline structures and by monitoring the
intensity of the strongest reflections during the metadynamics loop we can
easily detect when a phase transition occurs.  For a typical metadynamics
run, where the phase I transforms into phase III, we show in
Figure~\ref{fig:meta_1} the behavior of the box lengths and angles (top
panels) as well as the intensity of a selected peak of the structure
factor, relative to the strongest reflection.

\begin{figure*}[!htb]
  \begin{center}
    \includegraphics*[width=16cm]{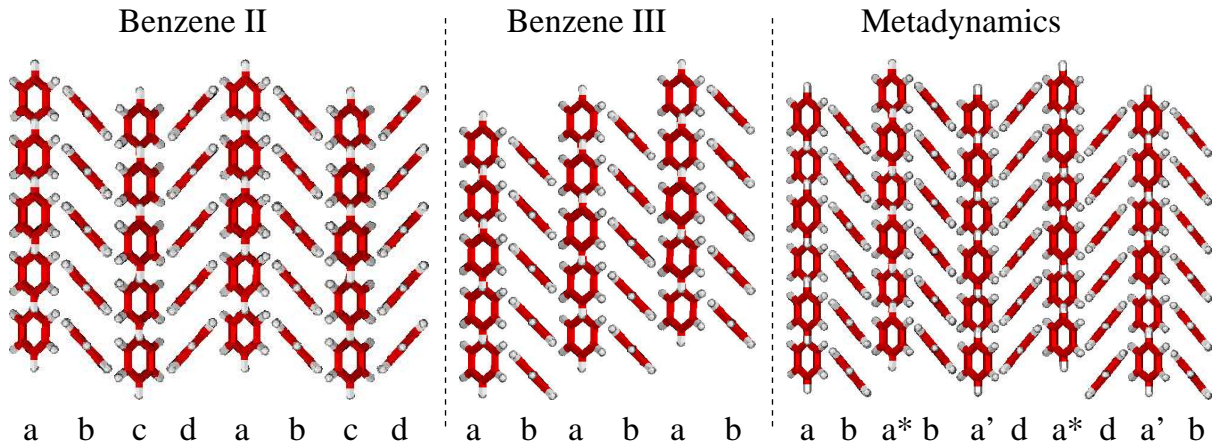}
  \end{center}
  \caption{Benzene: comparison of the molecular arrangement of phases
    II and III with that observed in a typical defective structure obtained
    during metadynamics. Note that benzene II has a typical {\it{a b c d}}
    plane stacking while benzene III has an {\it{a b }} stacking. The
    defective structure can be viewed either as a crystal of benzene II
    with two stacking faults (along the planes marked by a star) or as
    benzene III with two stacking faults (in the planes marked by a prime).
    \label{fig:fault}}
\end{figure*}

We tried tens of starting configurations with either of the three phases
and with different dimensions of the simulation cell, containing from 4 to
256 molecules. We observed that by increasing the cell dimension the number
of metasteps necessary to escape from the energy basin decreases.  At the
same time the formation of defects becomes easier and in the cells larger
than about hundred molecules we sometimes observe stacking faults.  We
stress, however, that we never observe ``point'' defects, such as vacancies
or misoriented molecules.  In particular, we could never observe a clean
transition between benzene I and II if the simulation cell was larger than
32 molecules. With larger cells benzene I always transforms either to
benzene III, which is lower in energy than benzene II, or to a defective
structure. In Figure~\ref{fig:fault} we show the plane stacking along the c
axis for benzene II and III and also for a typical defective configuration
obtained by metadynamics.

For comparison, we performed also a limited amount of Parrinello-Rahman
simulations, starting from phase I. We observed in this case only a
transition to phase III, but not to phase II. Moreover, no phase transition
is observed at pressures lower than 10~GPa. The results of metadynamics
simulations clearly demonstrate that the new method works very well also
for molecular crystals.

\section{Metadynamics using an internal order parameter}
\label{sec:metadyn-using-an}

The new method described in section \ref{sec:metadyn-using-simul} still
suffers from some limitations in common with the original Parrinello-Rahman
method.  For instance it is less effective for the study of phase
transitions for which the primary order parameter is an internal coordinate
instead of the cell edges. This is the case for phase transformations under
pressure described in terms of solid state chemical reactions such as the
2D \cite{C602D} and 3D \cite{C603D} polymerizations of C$_{60}$ or the
topochemical solid-state polymerizations of alkenes, alkynes and aromatic
hydrocarbons \cite{sissa6,poliP}. For instance, in the 2D polymerization of
C$_{60}$ the activation barrier for the [2+2] cycloaddition reaction is
overcome by a suitable deformation of the fullerenic cage which is not
induced by simply decreasing the intercage distances down to the density of
the 2D polymer \cite{CPL}.  In the perspective to address the study of
phase transformations in this class of materials, we have extended the
metadynamics scheme recently devised by Iannuzzi, Laio and Parrinello (ILP)
\cite{marcella} to constant-pressure MD simulations \cite{zipoli}.  The ILP
scheme can be dubbed {\sl reactive} molecular dynamics since suitably
defined reaction coordinates are introduced as dynamical variables.

By combining the ideas behind the Parrinello-Rahman and ILP methods we have
introduced a constant-pressure {\sl reactive} MD described by a Lagrangian
of the form
\begin{eqnarray}
\mathcal{L} &=&
\frac{1}{2} \sum_{i=1}^{N} m_i ({\bf\dot{s}}_i^t {\bf h}^t
{\bf h} {\bf\dot{s}}_i) - E(\{{\bf s}_i\},{\bf h}) +
\frac{1}{2} W_c Tr {\bf\dot{h}}^t {\bf\dot{h}} \nonumber \\ 
&-& p {\Omega} + \sum_{\alpha}\frac{1}{2}M_{\alpha}\dot{\eta}_{\alpha}^2
-\sum_{\alpha}\frac{1}{2}k_{\alpha}\bigl(\eta_{\alpha}(\{{\bf s}_i\},{\bf h})-
\eta_{\alpha}\bigr)^2 \nonumber \\ 
&-& V(t,\{\eta_\alpha\}),
\label{lagran}
\end{eqnarray}
where the first line is the Parrinello-Rahman Lagrangian \cite{pr_md} and
the second line is the ILP Lagrangian \cite{marcella}.  Here, ${\bf s}_i$
are scaled ionic coordinates, $\Omega$ is the cell volume, $p$ the external
pressure and $\eta_{\alpha}$ are collective variables as in the ILP scheme
\cite{marcella} with a fictitious kinetic energy and mass ($M_{\alpha}$). A
harmonic potential restrains the values of the collective coordinates
$\eta_{\alpha}(\{{\bf s}_i\},{\bf h})$ close to the corresponding dynamical
collective variables $\eta_{\alpha}$.  The values of $M_{\alpha}$ and
$k_{\alpha}$ control the time scale of the evolution of the collective
variables and are chosen according to the prescription given in Ref.
\cite{marcella}.  The collective coordinates $\eta_{\alpha}(\{{\bf
  s}_i\},{\bf h})$ are functions of the scaled ionic coordinates and of the
cell edges and should be able to discriminate between the initial and final
phases.  $E(\{{\bf s}_i\},{\bf h})$ is the total internal energy while
$V(\{\eta_{\alpha}\},t)$ is the history-dependent potential acting on the
collective variables and given by
\begin{equation}
V(t,\{\eta_{\alpha}\})=
\sum_{t'<t}W \prod_{\alpha}  e^{-\frac{\mid \eta_\alpha-\eta_\alpha^{t'}\mid^2}
{2\sigma_{\alpha}^2}},
\label{gauss} 
\end{equation}
where $W$ and $\sigma_{\alpha}$ are suitably chosen parameters as described
in Ref. \cite{hills}.  The equations of motion corresponding to the
Lagrangian (\ref{lagran}) are reported in Ref. \cite{zipoli}.

\subsection{Application to graphite}
\label{sec:application-graphite}

\begin{figure}[!htb]
\begin{center}
  \includegraphics[width=8.5cm]{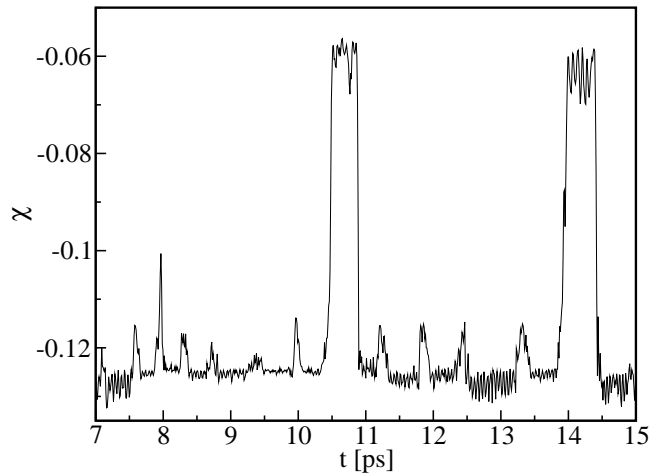}
\end{center}
\caption{Graphite: time evolution of the indicator $\chi$ (see
  Eq.~(\ref{sp2sp3}) in the text) which discriminates between the
  structures of graphite and diamond. Only the first part of the simulation
  is reported for sake of clarity.  Along the whole run 28 ps long, we have
  seen 15 jumps of $\chi$ similar to those reported in the figure which
  equally correspond to oscillations between graphite and
  diamond.\label{chi}}
\end{figure}

To demonstrate the validity of the new scheme described above we have
simulated the conversion of carbon from graphite to diamond under pressure
\cite{zipoli}.  This transformation can be seen as driven by an internal
order parameter such as the corrugation of the graphitic planes which leads
to the change of hybridization of carbon from sp$^2$ to sp$^3$.  As a
measure of the hybridization type of the carbon atoms, we have defined as
collective variable the coordination number of the atoms of a single
graphitic plane in the simulation cell with respect to the atoms of the two
neighboring planes, i.e.
\begin{equation}
\eta=\sum_{i\in plane}\sum_{j\notin plane} c_{ij},
\label{coord}
\end{equation}
with 
\begin{equation}
c_{ij}=\frac{ 1-(\frac{r_{ij}}{d})^6 }{ 1-(\frac{r_{ij}}{d})^{12} },
\end{equation}
where $r_{ij}$ is the distance between atoms $i$ and $j$ and $d=2.2$~\AA.
The graphite to diamond conversion has been already reproduced in the
ab-initio Parrinello-Rahman molecular dynamics simulations of Ref.
\cite{sissa7}, although at a pressure of 90 GPa, six times larger than the
experimental estimate (15 GPa, \cite{sissa7}), due to the aforementioned
limitations of the Parrinello-Rahman method.  In the present work graphite
is described by the tight-binding (TB) potential of Ref. \cite{wang}
supplemented by an empirical two-body van der Waals (vdW) interaction,
necessary to describe the interplanar distance in graphite \cite{zipoli}
\footnote{We verified that this modification of the TB potential has a
  negligible effect on the equation of state of the diamond phase.}.  This
model Hamiltonian describes with good accuracy the lattice parameters and
compressibility of graphite and diamond at equilibrium, but drastically
overestimates the compressibility of graphite at high pressure.  As a
consequence the theoretical transition pressure to diamond is as high as
129 GPa and the volume jump at the transition pressure is very small, the
volume being 4.70 $\rm\AA^3/atom$ for graphite and 4.60 $\rm\AA^3/atom$ for
diamond (cfr. the ab-initio EOS of graphite and diamond of Ref.
\cite{sissa7}).  Nevertheless this model graphite represents a good testing
case for the new simulation scheme. In fact, the small volume jump at the
theoretical transition pressure prevents the reduction of the activation
barrier by overpressurization and consequently the transition to diamond
does not take place in a Parrinello-Rahman simulation (70 ps long) even by
increasing the pressure up to 700 GPa and temperature up to 1000 K.
Conversely, within the new simulation scheme the transformation to diamond
occurs very close to the theoretical transition density.  We have performed
a metadynamics simulation of graphite at 15 GPa and 300 K with a supercell
containing 128 atoms initially arranged in the graphite structure with four
graphitic planes per cell in the ABAB (hexagonal) stacking \cite{sissa7}.
During the simulation run, 28 ps long, we have observed several (of the
order of 15) forward and backward transitions between graphite and diamond.
A clear monitoring of the phase transition is given by the time evolution
of the indicator $\chi$ which provides a sharp distinction between the
hybridization states of carbon atoms as
\begin{equation}
\chi=\frac{1}{N}\sum_i \frac{1}{n_i}\sum_{j>k}n_{ij}n_{ik}
cos^3\theta_{jik}
\label{sp2sp3} 
\end{equation}
where 
\begin{equation}
n_i=\sum_{j\ne i}n_{ij},
\end{equation}
\begin{equation}
n_{ij}=\frac{1}{1+e^{(r_{ij}-d)/\Delta}},
\end{equation}
with $d=1.8$ $\rm\AA$ and $\Delta=0.05$~\AA.  The index $i$ runs over all
the $N$ atoms of the simulation cell and $n_i$ is the coordination number
of atom $i$-th.  The index $k$ and $j$ run over atoms neighboring to atom
$i$-th and $\theta_{jik}$ is the angle subtended by the tern $jik$ whose
contribution to $\chi$ is weighted by the product of the partial
coordination number $n_{ik}$ and $n_{ij}$.  The cosine function in
(\ref{sp2sp3}) is able to discriminate between the hybridization sp$^2$ and
sp$^3$.  In fact the indicator $\chi$ has a value of -0.125 for graphite
and -0.06 for diamond at 15 GPa and 300 K. For the sake of clarity the time
evolution of $\chi$ is reported in Fig.~\ref{chi} only for the first part
of the run.  Two transitions from graphite to diamond are clearly
identified at 10.5 ps and at 14.0 ps. Similarly, other twelve
transformations can be identified in the other 12 ps of simulation.  The
Gaussian potential fills first the free-energy basin corresponding to
graphite and, once the system is driven to the new phase, the basin
corresponding to the final structure is then progressively filled.  Once
both basins are filled, the system is able to oscillate from one structure
to the other.  In agreement with the ab-initio Parrinello-Rahman simulation
of Ref.\cite{sissa7}, some of the structures observed during the
oscillations from graphite to diamond are a mixture of cubic and hexagonal
diamond.  The transformation at 10.5 ps (14.0 ps) starts at a volume of
4.45 $\rm\AA^3/atom$ (4.70 $\rm\AA^3/atom$), very close to the theoretical
transition volume of 4.70 $\rm\AA^3/atom$ obtained from the theoretical
equation of state of graphite and diamond.  These results demonstrate the
effectiveness of the metadynamics scheme presented above. The phase
transition occurs spontaneously at the theoretical transition density
whereas it does not take place in a Parrinello-Rahman simulation (within
the tight-binding model) even by overpressurizing the system up to five
times the theoretical transition pressure.  This metadynamics scheme would
be particularly suitable and probably superior to the one based on the
$\mathbf{h}$ matrix (as presented in section \ref{sec:metadyn-using-simul})
for the simulation of phase transitions described in terms of solid state
chemical reactions.

\section{Conclusions and outlook}
\label{sec:conclusions}

The methodology based on exploration of the free energy surface results in
a substantial improvement of simulation of structural transitions in
crystals. Using several inorganic and organic crystals as examples we
demonstrated its general applicability as well as heuristic value in
various cases where the plain constant-pressure simulations fail. The main
advantages that emerge can be summarized as follows. Starting from an
initial structure, the new approach is able to find a number of realistic
polymorphs, as demonstrated by the examples of zeolite and benzene
crystals. For the search for new structures, a classical force-field can be
used resulting in a simulation cost comparable to that of a standard
classical MD simulation.  The mechanism of transitions between different
phases can be understood at conditions close to those of experiment, as
shown by the example of graphite.

Clearly, there still remains a lot of room for improvement, including the
algorithm itself. As we see it, the major difficulty in the discrete
version of the algorithm described in section \ref{sec:metadyn-using-simul}
is the choice of parameters $W$ and $\delta h$.  One of possible
alternatives is to use also in this case a continuous version of
metadynamics with an adaptive mechanism for the choice of Gaussian
parameters (including anisotropic Gaussians), based on monitoring the cell
fluctuations and estimating the shape of the well. The work in this
direction is in progress.  Also, an accurate calculation of a transition
pressure at finite temperature still remains a challenge.  The ultimate
goal in this field should be a crystal structure prediction from
``scratch'', based only on the knowledge of the chemical composition of the
system. We believe that methods described in this paper represent a
substantial step forward towards this goal.

\end{document}